\documentclass[a4paper,fleqn]{cas-sc}

\usepackage[authoryear,longnamesfirst]{natbib}

\def\tsc#1{\csdef{#1}{\textsc{\lowercase{#1}}\xspace}}
\tsc{WGM}
\tsc{QE}

\newcommand{\be}{{\bf e}}
\newcommand{\bq}{{\bf q}}
\newcommand{\bR}{{\bf R}}
\newcommand{\bk}{{\bf k}}
\newcommand{\br}{{\bf r}}
\newcommand{\bp}{{\bf p}}
\newcommand{\bA}{{\bf A}}
\newcommand{\bB}{{\bf B}}

\newcommand{\rhobar}{\bar{\rho}}

\newcommand{\sgn}{\text{sgn}}

\begin{document}
\let\WriteBookmarks\relax
\def\floatpagepagefraction{1}
\def\textpagefraction{.001}

\shorttitle{From the Integer to the Fractional Quantum Hall Effect in Graphene}    

\shortauthors{M. O. Goerbig}  

\title [mode = title]{From the Integer to the Fractional Quantum Hall Effect in Graphene}  



%

\author[1]{Mark O. Goerbig}






\affiliation[1]{organization={Laboratoire de Physique des Solides},
            addressline={CNRS UMR 8502, Universit\'e Paris-Saclay}, 
            city={Orsay Cedex},
            postcode={F-91405}, 
            country={France}}




\begin{abstract}
The fractional quantum Hall effect is a very particular manifestation of electronic correlations in 
two-dimensional systems in a strong perpendicular magnetic field. It arises as a consequence of a strong Coulomb repulsion between electrons in the same Landau level that conspires 
with a particular chirality of the electronic states. This chirality is inherited from the classical cyclotron motion, \textit{i.e.} a particular sense of electronic rotation due
to the orientation of the magnetic field. The specificity of the FQHE in graphene consists of a four-fold spin-valley degeneracy inherited from the electronic bands in the vicinity 
of the Fermi level. The relevant Coulomb interaction respects this SU(4) symmetry, and one is therefore confronted with a generic four-component fractional quantum Hall effect, as
well as with other correlated four-component phases, such as spin-valley ferromagnetic states. The present article aims at a -- mainly theoretical -- discussion of these exotic 
phases, in comparison with experimental evidence for them. 
\end{abstract}



\maketitle

\section{Introduction}\label{sec:intro}

\begin{figure}[ht]
	\centering
		\includegraphics[width=0.98\textwidth]{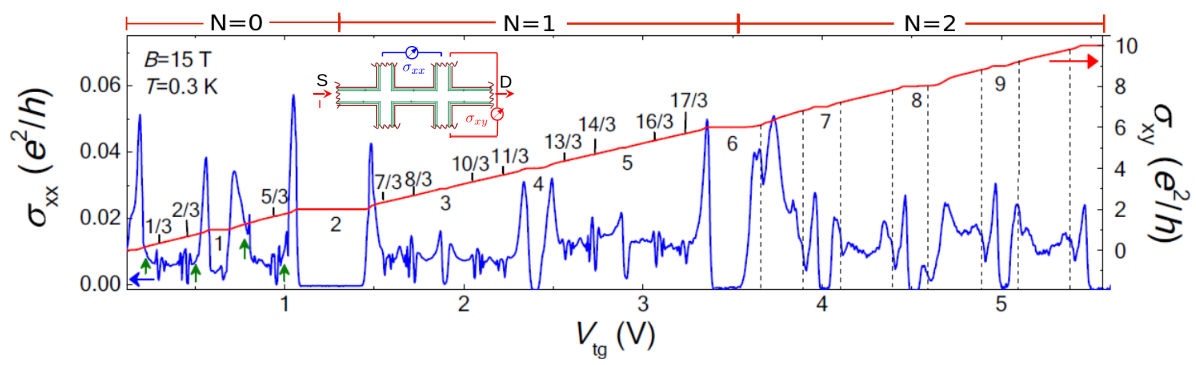}
	  \caption{Characteristic transport signatures of the quantum Hall effect in graphene [adapted from (\cite{rebeca2019})]. The Hall (blue) and the longitudinal (red) 
	  conductances are plotted as a function of the electronic density, which is proportional to the gate voltage $V_{tg}$, at fixed magnetic field ($B=15$ T) and temperature 
	  $T=0.3$ K. The inset shows the measurement setup in which a 
	  current is driven through the system via the contacts S (source) and D (drain). The longitudinal conductance is measured between the contacts connected by the blue line, 
	  while the Hall conductance is measured between the contacs connected by the red line. The numbers indicate the value $j$ of the plateau in the Hall conductance 
	  $\sigma_{xy}=R_H^{-1}=e^2 j/h$. One clearly distinguishes the integer plateaus of the IQHE (integer numbers below the Hall-conductance curve) from the FQHE plateaus at fractional 
	  values of $j$ (numbers above the Hall curve).}\label{fig01}
\end{figure}

The quantum Hall effect is a universal phenomenon of two-dimensional (2D) electronic systems submitted to a strong magnetic field. First discovered in 1980 by K. von 
Klitzing (\cite{klitzing1980}), who received the Nobel Prize in Physcs for this discovery in 1985, the effect consists of the appearance of plateaus in the Hall resistance accompanied
by a vanishing longitudinal resistance (for a sketch of the setup and a characteristic measurement, see Fig. \ref{fig01}, where the associated conductances are plotted). 
Most saliently, the plateaus in the Hall resistance 
indicate an extremely precise quantization, $R_H=h/e^2 j$, in terms of the inverse quantum of conductance $G_0=e^2/h$ and an integer number $j$. It is the manifestation of 
the so-called Landau quantization according to which 2D electrons in a strong magnetic field have a discrete energy spectrum with highly degenerate energy levels (Landau levels,
LLs). Soon after his discovery, 
D. Tsui and H. St\"ormer observed the fractional brother of the effect (\cite{Tsui1982}): instead of an integer in the case of the so-called \textit{integer quantum Hall effect} (IQHE), 
the number $j$ can be a fractional one, whence the name \textit{fractional
quantum Hall effect} (FQHE). It became soon clear that the effect requires strong electronic interactions that govern the behaviour of electrons inside a partially filled LL
and that force the electrons to form an incompressible electron liquid at particular values of the filling factor $\nu=n_\text{el}/n_B$, which denotes the filling of the LLs. It is 
the ratio between the electronic density $n_\text{el}$ and the density of flux quanta $n_B=eB/h$ threading the 2D system. A first approach to describe the 
FQHE was provided by R. Laughlin (\cite{Laughlin1983}), 
who proposed trial wave functions for the electron liquids at odd-integer filling factors $\nu=1/(2s+1)$, in terms of the integer $s$.
The in-built correlations of these wave functions turned out to be favourable
for the minimization of the mutual Coulomb repulsion between the electrons, and he shared the 1998 Nobel Prize in Physics with Tsui and St\"ormer for the discovery and description of the FQHE. One 
of the most exotic aspects of these electron liquids is certainly that of their quasiparticle excitations, which have fractional charge as it was first theoretically shown by Laughlin
(\cite{Laughlin1983}) and later experimentally proven in particular transport experiments. In the latter the so-called shot-noise was measured, which is proportional to the (fractional) charge 
(\cite{Picciotto,Glattli}). Associated with 
this fractional charge and the 2D nature of the system, these quasiparticles obey furthermore fractional statistics: they are anyons, \textit{i.e.} neither bosons nor fermions, and 
the identification of fractional statistics remains an important topic of today's research in condensed quantum matter (see also the chapter by M. Greiter and F. Wilczek), 
with recent breakthrough both in interference (\cite{Nakamura2020}) and transport measurements (\cite{Bartolomei}). 

While the observed states at $\nu=1/3$ and $\nu=1/5$ could indeed be understood in terms of Laughlin's wave function, a plethora of states at other partial filling factors has 
since been observed with the continuous increase of the sample quality in GaAs heterostructures. In order to account for those on a theoretical level, Laughlin's wave function 
has been generalized in several manners, the most noteworthy of which are Jain's composite fermion (CF) wave functions (\cite{Jain1989}) that account for the observation of states at 
$\nu=p/(2sp+1)$ -- the FQHE can then be seen as an IQHE of these CFs that populate $p$ CF LLs --, Halperin's wave functions (\cite{Halperin1983}) that account for states that are not 
or only partially spin-polarized, as well as the Pfaffian wave funtion (\cite{MR,GWW}), which describes a FQHE in half-filled LLs that has 
been observed at $\nu=5/2$ and $7/2$ (\cite{expMR}) in the first excited LL. Most interestingly, the quasiparticles of the Pfaffian state are a particular type of anyons: they 
are \textit{non-Abelian} anyons, in which case exchanges of more than two quasiparticles do not commute, and they are expected to play a relevant role in possible quantum
computation (\cite{kitaev}). (For more information about the FQHE in general, see the chapter by Z. Papi\'c and A. C. Balram.) 

The mechanical isolation of graphene in 2004 (\cite{Novoselov2004}) provided the condensed-matter-physics community with a novel 2D electronic system and has ever since been 
heavily studies because of its unexpected electronic properties and its remarkable relation with relativistic quantum mechanics. A milestone experiment in graphene research was the 
observation of a particular version of the IQHE in 2005 by two groups, that of A. K. Geim at Manchester University and that of P. Kim at Columbia University
(\cite{Novoselov2005,zhang2005}). The 
relevance of this discovery is twofold. First, it confirmed the above-mentioned universality of the effect since graphene is a largely different 2D electron system as compared to
the usual ones formed at semiconductor interfaces. Second, the particular filling-factor values $\nu=\pm 2(2n+1)$, at which the effect manifests itself, is a fingerprint of 
the relativistic quantum-mechanical behaviour of electrons in graphene that are described in terms of a Dirac rather than the usual Schr\"odinger equation of quantum mechanics. 
(For an introduction to the physics of graphene, see the chapter by E. McCann.) 

As compared to the original discoveries of the quantum Hall effects in the 1980ies, it took relatively longer to observe the FQHE in graphene, which was equally expected because
of the universality of the phenomenon. First experimental indications of a FQHE at $\nu=\pm 1/3$
were found in 2009 in freestanding graphene, where the original substrate on which the graphene 
sheet had been posed was etched away (\cite{grapheneFQHE1,grapheneFQHE2}). 
A true breakthrough was achieved in 2011 with the use of clean hexagonal boron-nitride (h-BN) flakes as a substrate that increased considerably
the electronic mobility in graphene. This increase of the mobility allowed for the observation of a large amount of FQHE states (\cite{Dean2011}), similarly to the effect in
the more conventional 2D electron systems in semiconductor heterostructures. 

The present article mainly aims at responding at the following question: what are the main differences with respect to the FQHE in conventional 2D electron systems? In analogy with
the IQHE, whose particular manifestation in graphene can be understood in terms of the relativistic nature of graphene's electrons, is there something such as a relativistic version of the 
FQHE? This was indeed a central issue in the theoretical discussions about the FQHE in graphene before its experimental discovery. After a short reminder of LL quantization in graphene 
and the IQHE in section \ref{sec:IQHE}, section \ref{sec:restr} introduces the basic properties of the electron dynamics in a single LL from a theoretical point of view. It is 
the basis for understanding the appearance of correlates phases, as it is discussed in section \ref{sec:phases}, not only the FQHE states (section \ref{ssec:FQHE}), but also
a particular form of spin-valley quantum Hall ferromagnetis the polarization in which does not only concern the usual electron spin but also another two-fold (valley) degree of
freedom (section \ref{ssec:QHFM}), and electron crystalline phases (section \ref{ssec:cryst}).

\section{Landau levels and the integer quantum Hall effect in graphene}\label{sec:IQHE}

Landau quantization denotes the formation of discrete levels into which the kinetic energy of 2D electrons is quenched in the presence of a strong perpendicular magnetic field.
It was first treated theoretically by L. Landau in 1930 for free non-relativistic particles with electric charge\footnote{Since we are interested, here, in electrons of 
negative charge, the following expressions are those for particles with charge $-e$, in terms of the elementary charge $e$, which we set to be positive.}
$-e$ that, in the absence of a magnetic field, 
have a quadratic dispersion relation, \textit{i.e.} a quadratic dependence of their energy on their wave vector $\bk$. In a typical condensed-matter situation, where one deals with
electrons in a periodic potential created by the atoms forming the crystal, this situation arises generically in the vicinity of the bottom of an electronic band that is 
described by a Bloch Hamiltonian $H(\bk)$. Elementary band theory shows that in the vicinity of the band bottom, where the Fermi level resides, \textit{e.g.} in the widely used
semiconductor GaAs, the Bloch Hamiltonian is $H(\bk)\simeq \hbar^2|\bk|^2/2m$, \textit{i.e.} precisely that of a free non-relativistic particle in terms of the band mass $m$. 
Landau quantization is then obtained by replacing the momentum $\bp=\hbar\bk$ by its gauge-invariant form $\bp + e\bA(\br)$, where $\bA(\br)$ is the vector potential at the 
position $\br$, which yields the magnetic field $\bB=\nabla\times\bA(\br)$. The usual quantum-mechanical non-commutativity of position and momentum, 
$[x,p_x]=[y,p_y]=i\hbar$, then gives the energy spectrum 
\begin{equation}
 E_n=\hbar\omega_C\left(n+\frac{1}{2}\right), \qquad n=0,1,2,...
\end{equation}
of the harmonic oscillator, in terms of the cyclotron frequency $\omega_C=eB/m$. In contrast to the usual one-dimensional harmonic oscillator, the energy levels (the LLs) 
are highly degenerate because the system remains 2D. This degeneracy is related to translation symmetry in the case of a homogeneous magnetic field. Even if this symmetry is,
strictly speaking, broken by the position-dependent vector potential entering in the Hamiltonian, translation symmetry is to some extent restored if we consider 
the electron dynamics from a (semi-)classical point of view. Indeed, the electrons are forced to perform a circular cyclotron motion around a position $\bR=(X,Y)$ -- the 
\textit{guiding centre} -- 
that is a constant of motion, in contrast to the momentum that constantly changes its direction. Most saliently, the two components of the guiding centre do not commute, 
$[X,Y]=il_B^2\sgn(B)$, in terms of the magnetic length $l_B=\sqrt{\hbar/e|B|}$ and the orientation of the magnetic field, $\sgn(B)=B/|B|$. This has two consequences: first, each 
quantum state in a LL has a particular \textit{chirality}, \textit{i.e.} a rotational sense imposed by the orientation of the magnetic field; second, the position 
of the guiding centre is not sharply defined, but submitted to a Heisenberg uncertaintly relation -- it is spread over a surface $2\pi l_B^2$
that can be viewed as the surface occupied by a quantum state in a particular LL. The total degeneracy $N_B$ of each LL is therefore given by the total surface $S$ devided by this minimal 
surface, $N_B=S/2\pi l_B^2=SB/(h/e)$, which is nothing other than the total flux threading the 2D system, in units of the flux quantum $h/e$. As a consequence of the fermionic 
nature of the electrons and Pauli's exclusion principle, the filling of the LLs is therefore determined by the filling factor 
\begin{equation}
 \nu= \frac{N_\text{el}}{N_B}=\frac{n_\text{el}}{n_B},
\end{equation}
in terms of the total number of electron $N_\text{el}=Sn_\text{el}$. 

The IQHE arises every time that an integer number of LLs is completely filled, \textit{i.e.} at $\nu=j$. In this case, the Fermi level resides between the 
LLs $n-1$ and $n$, and the system is a bulk insulator. Strictly speaking, $n$ should be an even integer 
due to the twofold spin degeneracy, but the latter is lifted either by the Zeeman effect or the formation of a ferromagnetic state due to exchange effects that favour a full spin 
polarization (see section \ref{ssec:QHFM}). The conductance is then governed by $j$ chiral ballistic edge channels the chirality of which is imposed by the orientation of the 
magnetic field, 
while additional electrons (or holes) that occupy adjacent LLs, if the filling factor is not precisely an integer, do not contribute to the
electric transport due to their localization by the bulk impurities. It is noteworth to mention that this is a particular form of the bulk-edge correspondance that makes the 
IQHE a prototype of a topological insulator, where the bulk is insulating but the edges are necessarily conducting. 

As already mentioned, the graphene is special because the band dispersion is not quadratic in the vicinity of the Fermi level. In charge-neutral (\textit{i.e.} undoped) graphene,
the Fermi level is situated at the two inequivalent points $K$ and $K'$ in the first Brillouin zone, where the valence band touches the conduction band with a dispersion that is linear in the wave vector.
This has two consequences. First, the low-energy spectrum in the vicinity of the Fermi level is \textit{twofold valley-degenerate}; second,  
the Bloch Hamiltonian must account for both conduction and valence bands on equal footing, in the form of the $2\times 2$-matrix form
\begin{equation}
 H(\bk) = \hbar v_F \left(\begin{array}{cc}
           0 & (k_x - ik_y)\\ (k_x+ik_y) & 0
          \end{array}\right).
\end{equation}
This is precisely the Dirac Hamiltonian of a massless particle in relativistic quantum mechanics, moving at the Fermi velocity $v_F$ instead
of the speed of light. Upon the same replacement of the wave vector by its gauge-invariant form as discussed above,
$\hbar \bk \rightarrow \bp + e\bA (\br)$, to account for the magnetic field, one obtains the LL spectrum
\begin{equation}
 E_n^G=\lambda \hbar \frac{v_F}{l_B} \sqrt{2n},
\end{equation}
where $\lambda=\pm$ denotes the two bands ($\lambda=+$ for the conduction and $\lambda=-$ for the valence band), and the ratio between the Fermi velocity and the magnetic length,
$v_F/l_B$,
plays the role of the cyclotron frequency here. It is clear from this spectrum that each LL in the conduction band has its counterpart in the valence band, apart from the $n=0$ LL
that is fixed at zero energy. 

Similarly to LLs in non-relativistic 2D electron systems with a quadratic band dispersion, the LLs of graphene are highly degenerate, and the number of states per LL is again given
by the number of flux quanta threading the 2D surface. Indeed, the argument invoked above -- that the centre of the cyclotron motion is a constant of motion and that there exists a
minimal surface occupied by any quantum state in a LL -- does not require a particular form of the electronic bands, but it is generally valid, regardless of the precise form of 
the LL spectrum.
However, the internal degrees of freedom are not the same. While in a conventional 2D electron system, one is simply confronted with the twofold spin degeneracy, in 
the absence of the Zeeman effect, graphene LLs are fourfold degenerate. In addition to the usual spin, one has a twofold valley degeneracy as a consequence of 
the above-mentioned two inequivalent points $K$ and $K'$ where the Fermi level crosses the electronic bands. The electrons are therefore formally described 
within an SU(4) symmetry, which arises as a consequence of the possibility to make a quantum-mechanical superposition of the orbital wave functions of each for the four spin-valley
combinations $|\uparrow, K\rangle$, $|\uparrow, K'\rangle$, $|\downarrow, K\rangle$, and $|\downarrow,K'\rangle$, where $s=\uparrow,\downarrow$ denotes the spin orientation. 
This particular SU(4) symmetry, which is respected to great extent in graphene LLs due to the smallness of the Zeeman effect and other valley-degeneracy lifting terms (\cite{goerbig2011}),
turns out to be of great relevance for the understanding not only of the FQHE in graphene, but also for the formation of exotic quantum-Hall ferromagnetic phases that are 
discussed in the following sections. If one discards these phases for the moment, one therefore expects an IQHE in graphene at filling factors that are spaced in units of four
due to the fourfold spin-valley degeneracy. This has indeed been observed experimentally (\cite{Novoselov2005,zhang2005}), but moreover the filling-factor series at which the 
IQHE occurs has an offset of two as compared to the IQHE in conventional 2D electron systems. This offset can be understood in the following manner. In contrast to non-relativistic
systems, where the absence of any charge carriers at $\nu=0$ leads to a completely empty lowest LL, charge neutrality in graphene ($\nu=0$) means that the zero-energy LL $n=0$ is
globally half-filled; there are as many electrons as holes that occupy the $n=0$ LL. Therefore, the Fermi level is situated inside this LL and not in the energy gap between two
adjacent LLs, as it is required for the occurance of the IQHE. The latter situation arises only at $\nu=2$, when the $n=0$ LL is completely filled, or at $\nu=-2$ when it is 
empty. In graphene, the IQHE thus occurs at the filling factors 
\begin{equation}
 \nu=\pm(4n+2)=\pm 2(2n+1),
\end{equation}
as it has been demonstrated experimentally (\cite{Novoselov2005,zhang2005}). This particular filling-factor sequence of the IQHE is a fingerprint
of the relativistic nature of the electrons in graphene.

\section{Electrons in a single Landau level}\label{sec:restr}

\begin{figure}[ht]
	\centering
		\includegraphics[width=0.78\textwidth]{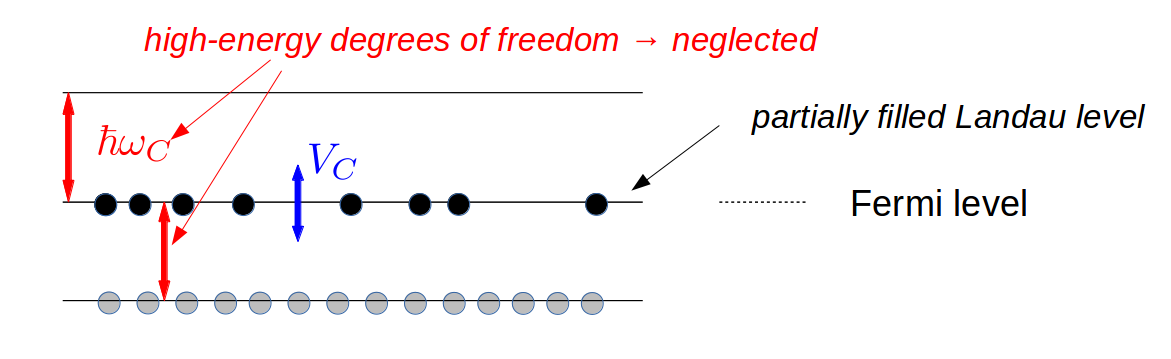}
	  \caption{Sketch of the low-energy model of electrons restricted to a single LL. The Fermi level is situated in the partially filled LL (black circles represent the occupied states), while the LLs below are completely 
	  filled (gray circles) and those above completely empty. Excitations across LLs (red arrows) are associated with an energy given by the LL separation $\hbar\omega_C$ and belong to the 
	  neglected high-energy degrees of freedom, while excitations in the same LL are gouverned by the Coulomb interaction $V_C$. }\label{fig02}
\end{figure}

In the preceding section, it was discussed that the IQHE arises as a consequence of Landau quantization every time that the Fermi level resides in between two adjacent LLs. As a 
consequence of the SU(2) spin symmetry in conventional 2D electron system, this is the case for $\nu=2(n+1)$, while the combination of (i) the relativistic nature of graphene 
electrons and (ii) their SU(4) spin-valley degeneracy yields an IQHE at $\nu=\pm 2(2n+1)$ in graphene. In the present section, we discuss which of these two particularities 
are inherited by the FQHE in graphene. Because the FQHE phases occur at fractionally filled LLs, one needs to restrict the electronic dynamics to a single LL: the 
low-energy excitations are now those inside the same LL whereas excitations across LLs can be treated as belonging to high-energy degrees of freedom of a characteristic 
energy given by the LL separation, \textit{i.e.} the energy scale $\hbar \omega_C$ in conventional 2D electron systems or $\hbar v_F/l_B$ in graphene. This is depicted in Fig.
\ref{fig02}. For the low-energy electronic excitations inside a single LL, the kinetic energy is therefore quenched, and the physical properties are governed by the 
Coulomb repulsion between the electrons.\footnote{There is a second natural energy scale that is set by the inevitable impurities in the 2D systems. However, since the FQHE and other
remarkable phases occur only in high-quality samples, the Coulomb repulsion is the leading energy scale and impurities should be treated perturbatively as a subordinate effect.
Their main role is then to localize the quasiparticles of the FQHE liquid and thus to fix the Hall resistance over a certain filling-factor range, whence the appearance of the 
Hall plateau. }

The low-energy electronic model finally appeals to the Coulomb interaction between the electronic densities $\rho_n(\br)$ restricted to a single LL $n$, 
\begin{equation}\label{eq:model0}
 H=\frac{1}{2}\int d^2r d^2r' \rho_n(\br) V(\br -\br')\rho_n(\br'),
\end{equation}
where $V(\br)=e^2/\epsilon |\br|$ is the usual Coulomb potential, which may eventually be screened due to a dielectric environment or the presence of gates. These
screening properties are generically taken into account by the dielectric function $\epsilon$. Let us now turn our attention in more detail to the LL-restricted densities
$\rho_n(\br)$, which are more conveniently described in reciprocal space, \textit{i.e.} by Fourier transformation,
\begin{equation}\label{eq:densop}
 \rho_n(\bq)=\int d^2r \rho_n(\br) e^{-i\bq\cdot\br} =\left\langle \sum_j e^{-i\bq\cdot\br_j}\right\rangle_n= F_n(\bq)\rhobar(\bq),
\end{equation}
where $\langle ...\rangle_n$ indicates the restriction to the $n$-th LL, of the Fourier components $\sum_j \exp(-i\bq\cdot\br_j)$ of the density operator in terms of the 
position operator $\br_j$ of the $j$-th particle. From a quantum-mechanical point of view, this restriction is necessary because the position operator has components in all
LLs, and restricting the electron dynamics to a single LL amounts to replacing the position operator precisely be that of the guiding centre $\bR_j$. As mentioned above, $\bR_j$ is a constant of motion and thus commutes with 
the kinetic Hamiltonian that gives rise to the LLs. The projected density operator in Eq. (\ref{eq:densop}) is thus implicitly defined as 
$\rhobar(\bq)=\sum_j\exp(-i\bq\cdot\bR_j)$.
The form factor $F_n(\bq)$ on the right-hand side of the expression (\ref{eq:densop}) takes into account the LL wave functions
that are system-dependent. In the case of the conventional 2D electron gaz (\textit{e.g.} in GaAs heterostructures), it is given in terms of Laguerre polynomials,
\begin{equation}
 F_n^\text{2DEG}(\bq)= L_n\left(\frac{q^2l_B^2}{2}\right)e^{-q^2l_B^2/4},
\end{equation}
while in graphene one needs to distinguish the $n=0$ LL from the other levels,
\begin{equation}\label{eq:LLFF}
 F_{n=0}^\text{G}(\bq)=e^{-q^2l_B^2/4}\qquad \text{and} \qquad F_{n\neq 0}^\text{G}(\bq)=\frac{1}{2}\left[ L_{n-1}\left(\frac{q^2l_B^2}{2}\right) +  L_n\left(\frac{q^2l_B^2}{2}\right)
 \right]e^{-q^2l_B^2/4}.
\end{equation}

The form factors can be absorbed into an \textit{effective interaction potential}
\begin{equation}\label{eq:effint}
 v_n(\bq) = \frac{2\pi e^2}{ \epsilon(\bq) |\bq|}\left[F_n(\bq)\right]^2,
\end{equation}
so that the electronic properties in a partially filled LL and thus the correlated phases responsible for the FQHE are governed by the model
\begin{equation}\label{eq:model}
 H=\frac{1}{2} \sum_\bq v_n(\bq) \rhobar(-\bq)\rhobar (\bq).  
\end{equation}
While this is a theoretical model that one would also start with for the description of electronic correlations in flat bands, the Hamiltonian (\ref{eq:model}) is quite special as a 
consequence of the magnetic field. Indeed, the magnetic field is still present via the guiding-centre operators, which enter into the expressions for the restricted density operators, 
and its quantum-mechanical commutation relations impose a particular chirality on the electronic states. They induce highly unusual commutation relations for the density operators 
(\cite{GMP})
\begin{equation}\label{eq:GMP}
 [\rhobar(\bq),\rhobar(\bq')]=2i \sgn(B) \sin\left(\frac{\bq\wedge\bq' l_B^2}{2}\right)\rhobar(\bq+\bq'),
\end{equation}
where $\bq\wedge\bq'=q_xq_y'-q_x'q_y$ is the 2D vector product. The chirality is hidden in the expression on the right-hand side of Eq. (\ref{eq:GMP}), the sign of which is determined by 
the orientation of the magnetic field in the $z$-direction perpendicular to the 2D plane. This chirality is inherited from that of the guiding-centre components, as discussed in 
the previous section. 

To summarize this theoretical section on the basic model encoding the exotic physical properties of the FQHE, the correlated phases require 
\begin{itemize}
 \item flat degenerate levels (or bands), that are the LLs here, \textit{i.e.} a quenched kinetic energy;
 \item repulsive interactions between the electronic density components inside this level, such as it is described by the Hamiltonians (\ref{eq:model0}) and (\ref{eq:model});
 \item a chirality of the quantum states that is encoded, here, in the commutation relation between the components of the guiding-centre operator, $[X,Y]=il_B^2\sgn(B)$, and 
 that yields a non-commutativity (\ref{eq:GMP}) to the Fourier components of the projected density operators;
 \item the precise form of the LL spectrum is irrelevant for the structure of this low-energy model, but it has an indirect effect via the effective interaction potential, which 
 accounts for the LL wave functions.
\end{itemize}
This summary is noteworthy since recent theoretical and numerical studies have shown that the FQHE phases are not restricted to partially filled LLs in 2D systems in a strong magnetic field, 
but they can be found also in partially filled flat electronic 2D bands with a non-zero Chern number (\cite{FCI}), in which case one speaks of \textit{fractional Chern insulators}.
Indeed, in this case the low-energy model for interacting electrons in a flat band has the same structure as that of 2D electrons in a single LL
if one uses an average Berry curvature that is homogeneous in reciprocal space
over the first Brillouin zone (\cite{parameswaran2011,goerbig2011}).
Finally, one should retain from this discussion that the difference between the FQHE in conventional 2D electron systems and that in graphene stems not from the particular form
of the LLs -- be they relativistic or non-relativistic -- but from (i) the different wave functions that yield slightly different effective interaction potentials and (ii) 
from the internal degrees of freedom, which turn out to be fourfold in the case of graphene because of the fourfold spin-valley degeneracy of its LLs. In contrast to the IQHE, there is
thus nothing such as a relativistic version of the FQHE in graphene.\footnote{Notice that a relativistic description has been adopted with great success in the description
of a half-filled LL that accounts naturally for the particle-hole symmetry manifest at this filling (\cite{Son}). However, this particular description is by no means inherited from the 
original LLs since it has been used also in the context of conventional 2D quantum Hall systems, such as in GaAs heterostructures.} The 
(effective) Coulomb interaction in Eq. (\ref{eq:effint}) respects this SU(4) symmetry; symmetry-breaking terms are suppressed algebraically in $a/l_B$, where $a=0.14$ nm is the characteristic
distance between nearest-neighbour carbon atoms in the graphene lattice, and the magnetic length is much larger at physically accesible magnetic fields, 
$l_B\simeq 24$ nm$/\sqrt{B[\text{T}]}$.

\section{Correlated electronic phases in partially filled Landau levels in graphene}\label{sec:phases}

Based on the model presented in the previous section, we now describe the main electronic phases encountered in partially filled Landau levels. Quite generally, one might 
already mention that due to the flatness of the Landau levels, spin- and valley-polarized states are favoured in order to minimize the exchange energy. This is reminiscent of 
the usual itinerant ferromagnetism, with the notable difference that in the case of LLs there is no cost in kinetic energy to fully polarize the electrons. This 
picture turns out to be the correct one in the description of quantum-Hall ferromagnetic (QHFM) states at integer filling factors other than the graphene sequence $\nu=\pm 2(2n+1)$,
and, because of the conceptually simpler description, we present the particular physical properties of these states in Sec. \ref{ssec:QHFM} before the FQHE. However, this picture
is more questionable in the theoretical description of the FQHE. As we discuss below, the \textit{exchange energy} might be outcast in several situations by the \textit{correlation energy}, 
which is intrinsically built into the trial wave functions such as the multi-component Halperin or Jain wave functions. In first exact-diagonalization studies that considered a 
completely polarized spin but a free valley polarization and that took into account the particular graphene LL form factors (\ref{eq:LLFF}),
the FQHE state at $\nu=1/3-2$ turned out to be fully polarized while that at $\nu=2/5-2$ is a valley singlet (\cite{AC2006}). The strong variation of the spin-valley polarization
of the various FQHE states has later been corroborated in theoretical studies that take into account the four-fold spin-valley degeneracy either in a four-component CF 
approach (\cite{TJ2007}) or in an approach that appeals to four-component generalizations of the Halperin wave function (\cite{GR2007}). 

From an experimental point of view, it is more difficult to obtain information about the spin-valley polarization of the various states. Signatures for the SU(4) nature of the 
$1/3$-family of the FQHE have been obtained in transport measurements of graphene on h-BN (\cite{Dean2011}) and later in compressibility measurements of the CF series
at $\nu=p/(2sp+1)$ in different subbranches of the $n=0$ LL in suspended graphene 
(\cite{Feldman2012}). However, they are not based on a direct measurement of the spin or valley polarization but on a relative comparison between the gaps and visibilities of 
the various states.

\subsection{Spin-valley quantum Hall ferromagnets}\label{ssec:QHFM}

The most straight-forward manifestation of the Coulomb repulsion is certainly the formation of QHFM states at integer fillings other than the characteristic
IQHE series $\nu=\pm 2(2n+1)$ in graphene. This is easily understood in analogy with ferromagnetism of itinerant electrons: in order to minimize the Coulomb repulsion, the 
orbital part of the overall $N$-electron wave function should be as antisymmetric as possible such that the probability to find two electrons at the same position is decreased. Because
of the fermionic global antisymmetry of the electronic wave function, the spin-valley part must therefore be symmetric, \textit{i.e.} maximally spin-valley polarized, that is 
precisely the hallmark of an SU(4) ferromagnetic state. Because of the flatness of the LL, this polarization is not accompanied by a cost in kinetic energy as for 
itinerant electrons in dispersive energy bands.

\begin{figure}[ht]
	\centering
		\includegraphics[width=0.48\textwidth]{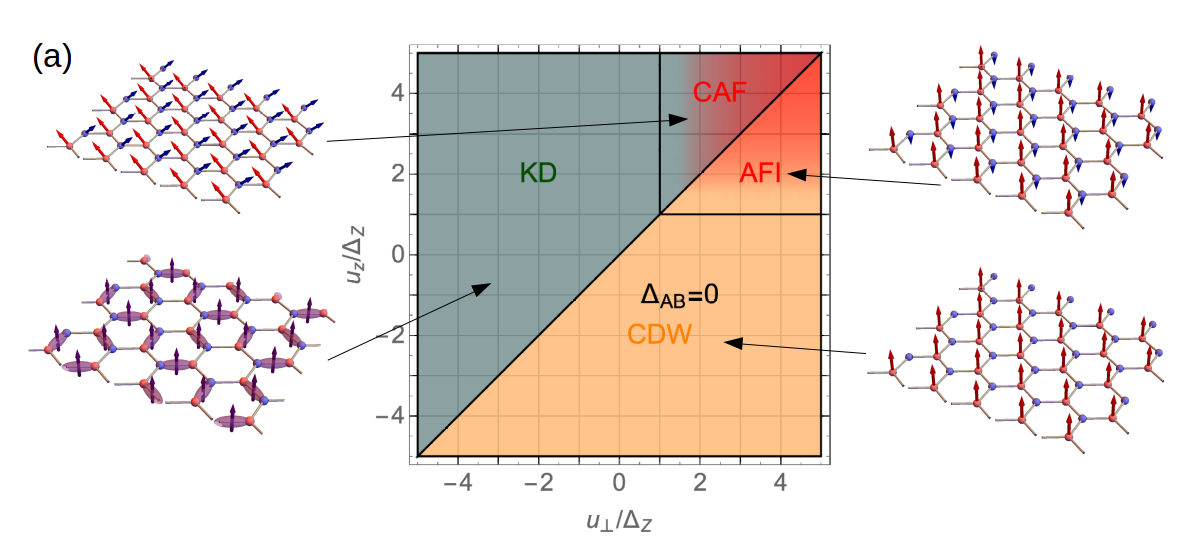}\includegraphics[width=0.48\textwidth]{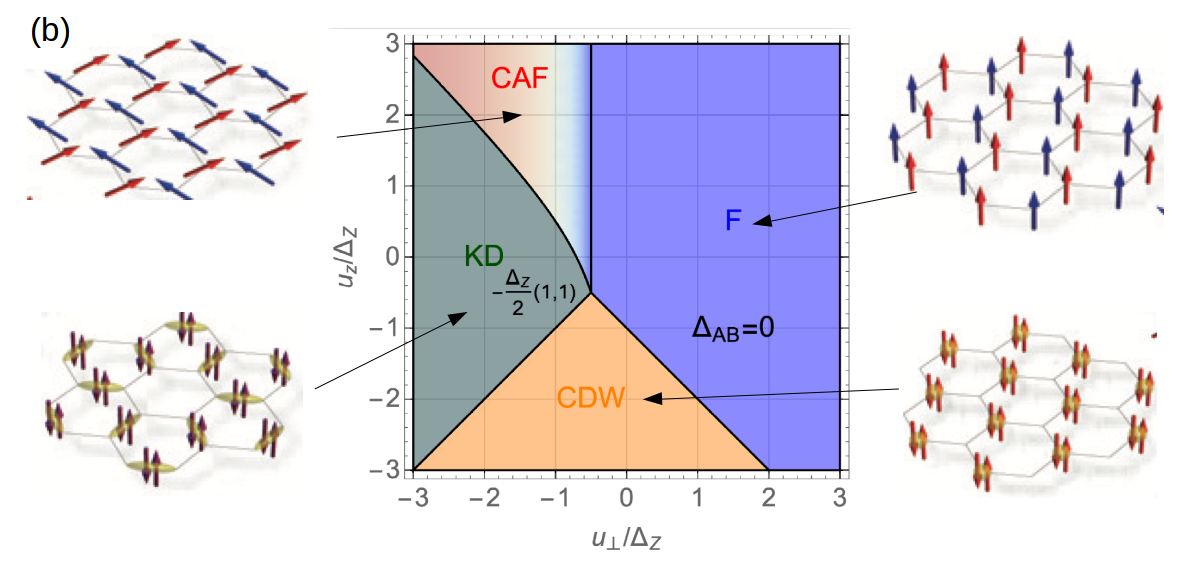}
	  \caption{Phase diagram for the QHFM phases at (a) $\nu=-1$ and (b) $\nu=0$, as a function of the parameters $u_z/\Delta_Z$ and $u_\perp/\Delta_Z$.
	  (a) At $\nu=-1$, the CDW and KD phases come along with a full spin polarization since the spin and
	  valley degrees of freedom are decoupled. On the contrary, the AFI and CAF phases show spin-valley entanglement. The sketches indicate the lattice-resolved spin densities in the 
	  different phases. (b) At $\nu=0$ the full valley polarization of the CDW and KD phases are spin-unpolarized, and the F phase with full spin polarization is therefore distinct 
	  from the former two phases. The CAF phase shows spin-valley entanglement.}\label{fig03}
\end{figure}

Let us illustrate the particular SU(4) QHFM states in the LL $n=0$. One needs to distinguish two cases, (i) $\nu=-1$, which is related by particle-hole symmetry to the 
case $\nu=+1$, and (ii) $\nu=0$ at charge neutrality. At $\nu=-1$, one of the spin-valley components $\alpha=|\uparrow,K\rangle$, $|\uparrow,K'\rangle$, $|\downarrow,K\rangle$
is completely filled, and the associated wave function reads
\begin{equation}\label{eq:QHFM1}
 \psi_{\nu=-1}=\prod_{k_\alpha<l_\alpha}^{N_B}\left(z_{k_\alpha}-z_{l_\alpha}\right)\exp\left(-\sum_{j_\alpha=1}^{N_B}|z_{j_\alpha}|^2/4l_B^2\right),
\end{equation}
in terms of the complex position $z_{j_\alpha}=x_{j_\alpha}-iy_{j_\alpha}$ of the $j_\alpha$-th electron in the component $\alpha$. Because of the SU(4) symmetry of the 
Coulomb interaction, there is no spin-valley component favoured over another one, and $\alpha$ could even represent an arbitrary quantum-mechanical superposition of the four 
components. Notice, however, that the SU(4) symmetry is eventually broken by interactions at the lattice scale (\cite{alicea,herbut,kharitonov2012}), electron-phonon 
interactions (\cite{elphon1,nomura09}) or
simply the Zeeman effect. Generally, these effects are associated with energy scales that are one or two orders of magnitude smaller than the leading SU(4)-symmetric Coulomb repulsion
(\cite{goerbig2011}), and we therefore consider the symmetric case first. Similarly, the QHFM wave function at charge neutrality ($\nu=0$) can be written as 
\begin{equation}\label{eq:QHFM0}
 \psi_{\nu=0}=\prod_{k_\alpha<l_\alpha}^{N_B}\left(z_{k_\alpha}-z_{l_\alpha}\right)\prod_{k_\beta<l_\beta}^{N_B}\left(z_{k_\beta}-z_{l_\beta}\right)
 \exp\left(-\sum_{j_\alpha=1}^{N_B}|z_{j_\alpha}|^2/4l_B^2-\sum_{j_\beta=1}^{N_B}|z_{j_\beta}|^2/4l_B^2\right),
\end{equation}
\textit{i.e.} the two components $\alpha$ and $\beta$ are now completely filled. As in the case $\nu=-1$, one may freely choose $\alpha$ and $\beta$ among the four components, 
including quantum-mechanical superpositions of them, as long as the components $\alpha$ and $\beta$ are orthogonal to one another. 

Both at $\nu=-1$ and $\nu=0$, the large variety of SU(4) polarizations is eventually fixed by the above-mentioned subleading symmetry-breaking terms, which are easily taken into account within 
a non-linear sigma model. In the conceptually simpler case $\nu=-1$, the energy due to these anisotropic terms can be written as (\cite{nomura09,Lian2017})
\begin{equation}\label{eq:anisot}
 E_A=\frac{N_B}{2}\left[u_z M_{P_z}^2 + u_\perp \left( M_{P_x}^2 + M_{P_y}^2\right)\right] - N_B\Delta_Z M_{S_z},
\end{equation}
where $M_{S_\mu}=\langle F|\sigma_\mu|F\rangle$ is the spin polarization of a general superposition $|F\rangle$ of the four different spin-valley components, in terms 
of the Pauli matrices $\sigma_\mu$ acting on the spin components of $|F\rangle$. Similarly, $M_{P_\mu}=\langle F|\tau_\mu|F\rangle$ is the valley polarization of $|F\rangle$ 
in terms of the Pauli matrices $\tau_\mu$ that now act on the valley components of $|F\rangle$. The usual Zeeman effect is taken into account by the energy scale $\Delta_Z$, while 
$u_z$ and $u_\perp$ act on the valley components. As mentioned above, they account for either lattice-scale interactions (\cite{alicea,herbut,kharitonov2012} or electron-phonon coupling. 

The resulting
phase diagram is shown in Fig. \ref{fig03}(a), as a function of $u_z/\Delta_Z$ and $u_\perp/\Delta_Z$, and can be understood in the following manner. Apart from the 
\textit{canted antiferromagnetic} (CAF) and the \textit{antiferrimagnetic} (AFI) phases, the spin and valley polarizations are fully decoupled. Indeed, at $\nu=-1$, the system 
can be both fully spin- and fully valley-polarized, \textit{e.g.} if all electrons occupy the state $|F\rangle=|\uparrow, K\rangle$. This is the case in the \textit{charge-density-wave}
(CDW) phase that is stabilized if $u_z<u_\perp<\Delta_Z$. The particular sublattice occupation, where only one sublattice is occupied, is a consequence of the particular nature
of the one-particle wave functions in the graphene $n=0$ LL, where sublattice and valley are identical: electrons of the valley $K$ have a non-zero wave function only on one 
sublattice, while electrons of the valley $K'$ solely occupy the other sublattice. Similarly, for $u_\perp<u_z<\Delta_Z$, the valley polarization is preferentially in a superposition
of the valleys $K$ and $K'$, along with a full spin polarization in the $z$-direction. This superposition manifests itself in a particular lattice pattern in the form of a 
\textit{kekul\'e distortion} (KD). 

The phases CAF and AFI merit special attention. They are specific SU(4) ferromagnetic phases that have no counterpart in the usual SU(2) ferromagnetism since they are driven 
by an entanglement between the spin and valley degrees of freedom. Consider for example the AFI phase, which becomes antiferromagnetic in the limit of a vanishing Zeeman gap, 
$\Delta_Z\rightarrow 0$. From the above-mentioned perspective, it is perfectly admitted to construct an equal-weight superposition of the states $|\uparrow, K\rangle$ and 
$|\downarrow,K'\rangle$ of all electrons. In this case, the SU(4) QHFM consists only of spin-$\uparrow$ electron in the $K$ valley, \textit{i.e.} on one sublattice, while the 
other sublattice (associated with the valley $K'$) only hosts spin-$\downarrow$ electrons. This would commonly be referred to as an \textit{antiferromagnetic} state, but one must 
insist that it is an SU(4) \textit{ferromagnetic} phase which can be transformed into one of the other QHFM phases by a global SU(4) rotation. This state is favoured at positive 
values of $u_z$ and $u_\perp$, which prefer to have a vanishing valley polarization in all directions, as one may see from Eq. (\ref{eq:anisot}). As a matter of fact, due to the spin-valley entanglement, a vanishing valley polarization comes along with a vanishing spin polarization (\cite{Doucot2008}). However, a fully vanishing spin polarization can be achieved only in the limit of zero Zeeman effect, which otherwise prefers a non-zero value of the spin polarization in the $z$-direction, whence the term 
\textit{antiferrimagnetic}. Similarly the CAF phase has a net spin 
polarization out of plane to account for the Zeeman effect, but an inplane antiferromagnetic pattern. 

The phase diagram at $\nu=0$ has similar phases (\cite{kharitonov2012}), but one needs to take into account the fact that now two components are fully occupied by electrons, as 
stipulated by the wave function (\ref{eq:QHFM0}). In this case, one cannot polarize fully both the spin and the valley, \textit{i.e.} one now needs to make a distinction 
between the spin-polarized \textit{ferromagnetic} (F) phase and the valley-polarized phases CDW and KD. The F phase comes along with a zero valley polarization while the 
CDW and KD phases are accompanied by a vanishing spin polarization. Again, it is possible to profit from spin-valley entanglement via special superpositions, such as in the CAF phase.
The phase diagram (\cite{kharitonov2012,Atteia2021a}) is shown in Fig. \ref{fig03}(b) for the same parameters $u_z$, $u_\perp$ and $\Delta_Z$. The precise values of these parameters 
are yet unknown and they are likely to depend on the concrete physical system, \textit{e.g.} the (dielectric) environment of the graphene sheet or its amount of disorder. Recent
scanning-tunneling-spectroscopic measurements have found evidence for all four phases at $\nu=0$ (\cite{yazdani2022,Coissard2021}), for graphene on different substrates.

\subsubsection{Magnons}

\begin{figure}[ht]
	\centering
		\includegraphics[width=0.48\textwidth]{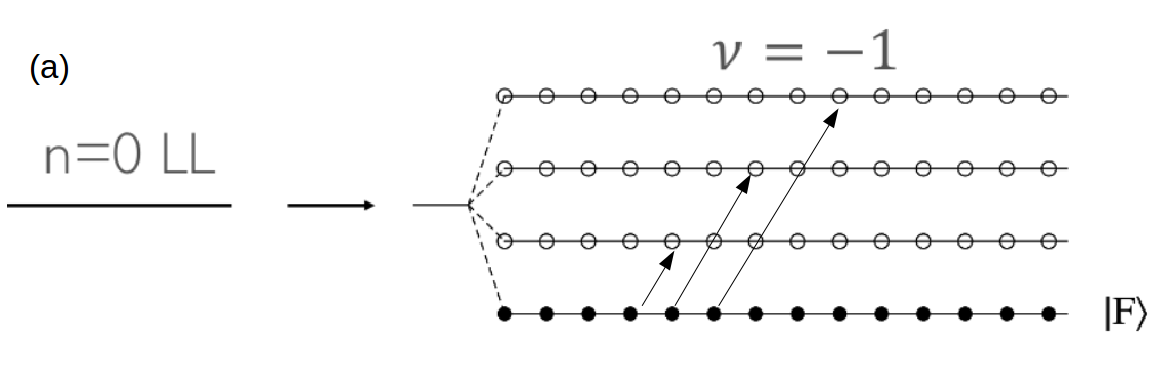}\includegraphics[width=0.48\textwidth]{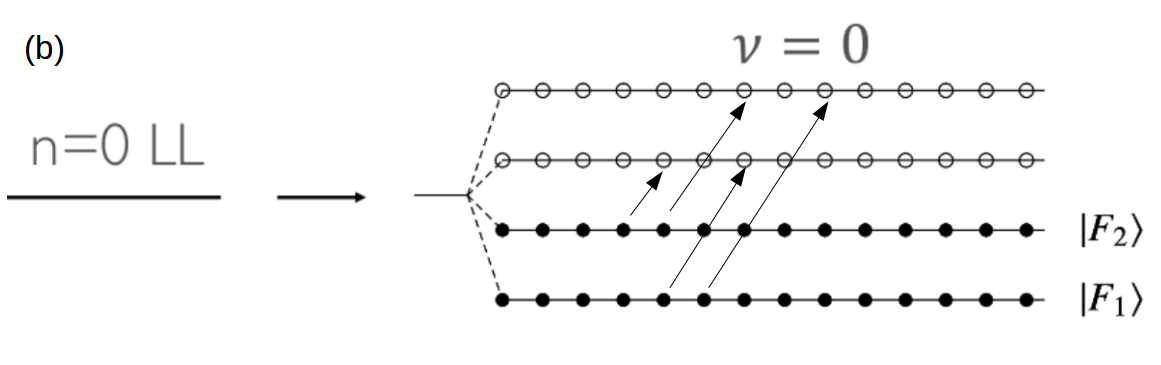}
	  \caption{Generic magnons at (a) $\nu=-1$ and (b) $\nu=0$. The $n=0$ LL hosts four spin-valley copies that are separated for illustration reasons while they remain 
	  degenerate in energy in the absence of explicit SU(4)-symmetry-breaking terms. At $\nu=-1$ (and its particle-hole-symmetric situation at $\nu=1$) one of the spin-valley
	  components is completely occupied and one obtains three magnon modes. At $\nu=0$, two components are filled and one has four magnon modes. 
	  }\label{fig04}
\end{figure}

As a consequence of the spontaneously broken SU(4) symmetry in the formation of a QHFM, special low-energy collective excitations, called \textit{Goldstone modes}, occur 
if one perturbs slightly the ground states. They are nothing other than generalized spin waves (magnons) that happen again to be special in graphene as a consequence of the relatively
large SU(4) symmetry. In addition to the usual spin waves one has valley waves and mixed spin-valley magnons. This can be seen most easily again at $\nu=-1$, for which the 
situation is depicted in Fig. \ref{fig04}(a). Indeed, a simple counting of the possible modes indicates that there are three magnon branches. In the case of an SU(4)-symmetric
interaction all three branches have the same dispersion
\begin{equation}
 E_\bq= 2\sum_\bk v_n(\bk) \sin^2\left(\frac{\bq\wedge \bk l_B^2}{2}\right), 
\end{equation}
in terms of the effective interaction (\ref{eq:effint}) and the pair wave vector $\bq$. Indeed, a magnon can be viewed as a superposition at a specific wave vector $\bq$
of a hole in the component $\alpha$, which is fully occupied in the QHFM state $|F\rangle$, and an electron in the originally empty component $\beta$. 
Since the superposition is charge-neutral,
the pair wave vector $\bq=\Delta \bR\times \be_z/l_B^2$, which is related to the spatial separation $\Delta \bR=\bR -\bR'$ between the guiding centre of the electron $\bR$ and that
of the hole $\bR'$ and the orientation of the magnetic field $\be_z=\bB/|B|$, remains a constant of motion even in the presence of a magnetic field, contrary to the wave vectors
of the original (charged) constituents of the pair. 
In the central $n=0$ LL, this yields the magnon dispersion (\cite{KH,yang2006,doretto})
\begin{equation}\label{eq:dispMagn}
E_\bq= \sqrt{\frac{\pi}{2}}\frac{e^2}{\epsilon l_B}\left[ 1- e^{-q^2l_B^2/4}I_0\left(\frac{q^2l_B^2}{2}\right)\right],
\end{equation}
in terms of the modified Bessel function $I_0(x)$. The two limits for small and large wave vectors are readily understood; at small wave vectors, one retrieves the usual quadratic
\begin{equation}
 E_{q\rightarrow 0}\simeq \frac{\rho_s}{2}q^2l_B^2
\end{equation}
behaviour of the Goldstone mode, in terms of the spin stiffness 
\begin{equation}
 \rho_s=\frac{1}{16\sqrt{2\pi}}\frac{e^2}{\epsilon l_B},
\end{equation}
while at large wave vectors, the dispersion saturates at twice the exchange gap $E_X=\sqrt{\pi/8}e^2/\epsilon l_B$,
\begin{equation}
 E_{ql_B\gg 1}\simeq  \left[2\sqrt{\frac{\pi}{8}}- \frac{1}{ql_B^2}\right]\frac{e^2}{\epsilon l_B}.
\end{equation}
Because the exchange gap $E_X$ can be viewed as the energy to create a quasiparticle or a quasihole with a reversed spin, $2E_X$ is nothing other than the particle-hole dissociation
energy, while at finite wave vectors with $ql_B>1$ one needs to take into account the binding energy between the electron and the hole, which is given by the Coulomb energy 
$E_\text{Coul}=e^2/\epsilon |\Delta \bR|=e^2/\epsilon ql_B^2$ in the second term of the dispersion relation. 

Similarly to the QHFM ground states, the magnon spectrum becomes modified when the anisotropic terms (\ref{eq:anisot}) are taken into account. Because the degeneracy between the 
SU(4) QHFM states is lifted, the magnons depend on the precise form of the chosen state. Furthermore, the magnons become generally gapped, \textit{e.g.} by the Zeeman term 
for the spin magnons and the terms $u_z$ and $u_\perp$ for the valley magnons (\cite{Atteia2021b}). Most saliently, the magnon dispersion may even change its wave-vector 
dependence, \textit{e.g.} in the KD phase. Indeed, in this phase the energy functional, even in the presence of the anisotropic terms, remains U(1)-symmetric, while this symmetry
is spontaneously broken by the intervalley coherence in the KD phase. This yields a particular gapless magnon with a dispersion that is linear in the wave vector. 

Finally, the above energy arguments remain valid at charge neutrality ($\nu=0$), but the counting of the possible modes is changed. Because there are now two fully occupied 
spin-valley components (and two empty ones), one finds four possible magnon modes
(\cite{yang2006,Lambert2013,Wu2015,Wei2021}). However, for an SU(4)-symmetric Coulomb interaction, the magnon dispersion is still given by 
Eq. (\ref{eq:dispMagn}), and again some of the different magnon modes become gapped in the presence of an explicit spin-valley anisotropy (\ref{eq:anisot}).

\subsubsection{Skyrmions}

\begin{figure}[ht]
	\centering
		\includegraphics[width=0.68\textwidth]{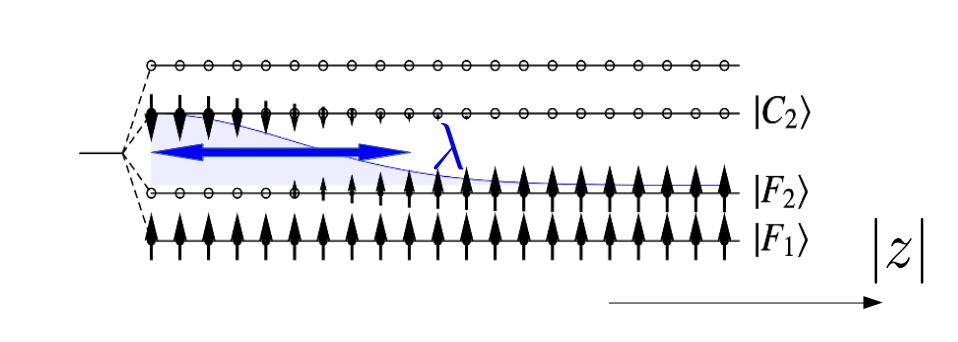}
	  \caption{Generic skyrmion at $\nu=0$. The sublevels $|F_1\rangle$ and $|F_2\rangle$ are occupied in the QHFM ground state. The skyrmion consist of an electron that is 
	  promoted from the spin-valley component $|F_2\rangle$ to the initially unoccupied component $|C_2\rangle$ at the skyrmion core $z=0$. Slightly away from the core, the electrons are 
	  in a superposition of $|C_2\rangle$ and $|F_2\rangle$, the relative weight of which changes upon increase of $|z|$. The typical size $|\lambda|$ indicates the equal-weight 
	  superposition
	  of the two components, while at $|z|\rightarrow \infty$ the superposition converges to the QHFM background $|F_2\rangle$. At all distances, $|F_1\rangle$ remains unchanged
	  as well as the unoccupied spectator component $|C_1\rangle$. }\label{fig05}
\end{figure}

A very particular type of topological objects are formed on top of the QHFM when additional charges are added to the ground states at $\nu=-1$ or $\nu=0$. Consider an electron added
to a perfect QHFM state at $\nu=0$. Naturally, this additional electron must occupy a spin-valley component that was initially completely empty. From an exchange-energy point 
of view it is however much more favourable that this \textit{spin-valley flip} does not only concern the additional electron, but the latter encourages electrons in its vicinity 
to do likewise such as to minimize the overal exchange energy. This is depicted in Fig. \ref{fig05} for a generic situation at $\nu=0$. The skyrmion is a texture that appeals to 
one of the two originally occupied spin-valley components (here $|F_2\rangle$) and one of the unoccupied ones (here $|C_2\rangle$). As in the QHFM ground state, these components 
can be any type of superposition of the original spin-valley components, but all states, the occupied ones $|F_1\rangle$ and $|F_2\rangle$ as well as the unoccupied ones 
$|C_1\rangle$ and $|C_2\rangle$, must be mutually orthogonal. While the QHFM ground state fixes the states $|F_1\rangle$ and $|F_2\rangle$, there is yet a certain choice for the 
spinor $|C_2\rangle$ that describes the spin-valley polarization at the skyrmion core. Indeed, the skyrmion can be generically described by the field 
\begin{equation}
 |Z(x+iy)\rangle= \frac{1}{\sqrt{1+(x^2+y^2)/|\lambda|^2}}\left( |C_2\rangle + \frac{x+iy}{\lambda} |F_2\rangle\right),
\end{equation}
where $|\lambda|$ denotes the typical skyrmion size. As for the QHFM states, the lowest-energy skyrmion that is realized in the system depends on the SU(4)-symmetry-breaking 
terms (\ref{eq:anisot}), and both at $\nu=-1$ (\cite{Lian2016,Lian2017}) and at $\nu=0$ (\cite{Atteia2021a}), one finds a whole \textit{skyrmion zoo} that is not presented here. 
However, it needs to be underlined that very recently there has been first spectroscopic evidence for the presence of skyrmions in graphene at $\nu=0$ (\cite{yazdani2022}),
in scanning-tunneling spectroscopy. While the QHFM is likely to be a KD phase, the spin-valley pattern at the skyrmion core is reminiscent of the CAF phase. Both phases have a 
particular signature in the sublattice occupation that has been precisely measured in the experiment.

\subsection{Fractional quantum Hall states}\label{ssec:FQHE}

In order to describe the four-component FQHE in graphene, one may start with a four-component generalization of Halperin's wave function (\cite{GR2007}),
\begin{equation}\label{eq:halperin}
 \psi_{m_1,...,m_4;n_{\alpha\beta}}=\phi_{m_1,...,m_4}^L\phi_{n_{\alpha\beta}}^\text{inter}\exp\left(-\sum_{\alpha=1}^{4}\sum_{j_\alpha=1}^{N_\alpha}|z_{j_\alpha}|^2/4l_B^2\right),
\end{equation}
where 
\begin{equation}
 \phi_{m_1,...,m_4}^L=\prod_{\alpha=1}^4\prod_{k_\alpha<l_\alpha}^{N_\alpha}\left(z_{k_\alpha}-z_{l_\alpha}\right)^{m_\alpha}
\end{equation}
is the product of 4 Laughlin wave functions for each of the four components $\alpha,\beta=|\uparrow,K\rangle$, $|\uparrow,K'\rangle$, $|\downarrow,K\rangle$, and 
$|\downarrow,K'\rangle$. In the absence of inter-component correlations, which are taken
into account in the term 
\begin{equation}
 \phi_{n_{\alpha\beta}}^\text{inter}=\prod_{\alpha<\beta}^4\prod_{k_\alpha}^{N_\alpha}\prod_{l_\beta}^{N_\beta}\left(z_{k_\alpha}-z_{l_\beta}\right)^{n_{\alpha\beta}},
\end{equation}
the exponents $m_\alpha$ are directly related to the component filling factors $\nu_\alpha=N_\alpha/N_B=1/m_\alpha$, which enter the total filling factor 
\begin{equation}
 \nu=\sum_\alpha \nu_\alpha - 2,
\end{equation}
in terms of the number of electrons per component $N_\alpha$ and the total number of electron $N_\text{el}$. The offset of 2 in the expression for the total filling factor is due to the 
fact that the $n=0$ LL in graphene is completely empty (for $\nu_\alpha=0$) at $\nu=-2$ because charge neutrality ($\nu=0$) corresponds to a globally half-filled $n=0$ LL, as mentioned
in Sec. \ref{sec:IQHE}. In general, however, the the total filling factor for a four-component Halperin wave function depends not only on the exponents $m_\alpha$ but also on
the inter-component correlations $n_{\alpha\beta}$, via the relation 
\begin{equation}
 1= m_\alpha \nu_\alpha + \sum_{\beta\neq \alpha} n_{\alpha\beta}\nu_\beta.
\end{equation}

Notice that the exponents $m_\alpha$ and $n_{\alpha\beta}$ induce correlations between the electrons. While the Pauli 
principle (\textit{i.e.} the antisymmetry of the fermionic wave functions) only requires $m_\alpha=1$ and does not constrain the inter-component correlations, the Halperin wave 
function (\ref{eq:halperin}) indicates that the electronic density of an electron of type $\alpha$ approching another one of the same type is suppressed as $\sim r^{2m_\alpha}$, 
and it is suppressed as $\sim r^{2n_{\alpha\beta}}$ if an electron of type $\alpha$ approaches one of type $\beta$. These correlations are therefore globally favourable 
for the repulsive Coulomb interaction. However, not all combinations of exponents are eligible for physically relevant wave functions: if the inter-component correlations become 
stronger than the intra-component ones, the Halperin wave function no longer describes a spatially homogeneous electronic system but electrons in different components have a 
tendency to undergo a phase separation (\cite{deGail2008}).
The four-component CF wave functions are more complex to write down, but the basic idea is to first rewrite a multi-component wave function 
of Halperin's type as a product 
\begin{equation}
 \psi = \phi_{\nu^*=1} \phi_\text{SU(4)}^{2s}\exp\left(-\sum_{\alpha=1}^{4}\sum_{j_\alpha=1}^{N_\alpha}|z_{j_\alpha}|^2/4l_B^2\right),
\end{equation}
as a vortex part $ \phi_\text{SU(4)}^{2s}$, and a wave function $\phi_{\nu^*=1}$ at a putative filling factor $\nu^*=1$ for electrons of one, several or all components. In a 
second step, this wave function is then replaced by one for $\nu^*=p$ LLs that are called CF LLs (\cite{TJ2007}). Since this wave function has non-analytical components, \textit{i.e.}
components in LLs other than $n=0$, the new wave function needs to be ultimately projected into this LL.

\subsubsection{The $1/3$-family of fractional states in graphene}

The generalized Halperin wave functions (\ref{eq:halperin}) may used as a first approach
to understand the particular form of the $1/3$-family of fractional states, which has been 
observed in graphene on a h-BN substrate and that has been interpreted as a manifestation of the particular SU(4) symmetry of the FQHE (\cite{Dean2011}). Indeed, if the 
fourfold spin-valley degeneracy were lifted by external perturbations such as the Zeeman effect, which favours a particular spin orientation, all four subbranches of the 
LLs would be well separated in energy. In this case, one may argue that the electrons in each of the subbranches form correlated states independently of the other (completely filled
or completely empty) subbranches. The latter would be electronically inert similarly to the other LLs that are remote in energy from the partially filled one, as we have discussed 
in Sec. \ref{sec:restr} where we constructed the low-energy model of electrons restricted to a single LL. In this case, one would expect to observe all spin-valley copies of the 
$1/3$ Laughlin state at $\nu=\pm 5/3,\pm 4/3, \pm 2/3,\pm 1/3$ in the $n=0$ LL. However, this was not observed in the experiment (\cite{Dean2011}), where the $\pm 5/3$-states were almost absent in the
measurement. In later experiments, at larger magnetic fields, the latter states became more pronounced (\cite{Polshyn2018}), in line with the intuitive expecations that at higher magnetic field the 
subbanches are better separated in energy (due to a $\propto B$ scaling with the magnetic field, in contrast to the expected $\sqrt{B}$-scaling of the Coulomb interaction scale
$e^2/\epsilon l_B$) such that a one-component treatment in terms of Laughlin wave functions becomes more reliable. 

In order to describe the $-5/3$-state (and its particle-hole symmetric one at $\nu=+5/3$), one can use the Halperin wave function (\ref{eq:halperin}), in which one sets 
all exponents $\mu_\alpha=n_{\alpha\beta}=3$. In this case, one may interpret the state as an overall Laughlin wave function that is built up from all electrons regardless of the 
spin-valley components they occupy. In spite of its similarity with the usual one-component Laughlin state, one needs to emphasize that it is an intrinsic multi-component state
with an internal SU(4) spin-valley symmetry, which allows one to distribute the electrons freely over the different spin-valley components. This freedom has consequences in the 
excitation spectrum in the form of generalized spin waves, similarly to the QHFM states at $\nu=\pm 1$ (see Sec. \ref{ssec:QHFM}): 
there are altogether three modes, a spin magnon, a valley magnon and a mixed spin-valley mode,
while the usual one-component Laughlin wave function would not have such excitations. The presence of these modes is likely to be at the origin of the reduced robustness of the $\pm 5/3$ states as compared to the other ones of the $1/3$ family. 

The states at $\pm 1/3$ can also be understood in terms of the generalized Halperin wave functions (\cite{Papic2010}). For illustration, consider a prominent Zeeman effect that favours a 
spin polarization over a valley polarization.\footnote{The argument does not depend on this choice, however, and one may exchange spin and valley or any two components.} 
In this case, we have the exponents $m_\alpha=1$ for $\alpha=|\uparrow,K\rangle$ and $|\uparrow,K'\rangle$, and $m_\alpha=3$ for $\alpha=|\downarrow,K\rangle$ and 
$|\downarrow,K'\rangle$. Furthermore, there are no correlations between $|\uparrow,K\rangle$ and $|\uparrow,K'\rangle$ such that the corresponding exponent is $n_{12}=0$,
while it is $n_{34}=3$ for the correlations between $|\downarrow,K\rangle$ and $|\downarrow,K'\rangle$. Finally, there are no correlations between states of different spin
orientation, $n_{13}=n_{14}=n_{23}=n_{24}=0$. Similarly to the $\pm 5/3$-states, this state has a residual symmetry in the $s=\downarrow$-spin branch since one can freely distribute the 
electrons of this spin orientation over the two valleys, \textit{i.e.} one is confronted with a residual SU(2) symmetry that is spontaneously broken by the formation of the 
state and that supports thus a valley magnon. In addition to the number of generalized spin waves, there is an important difference between the Halperin states at $\nu=\pm 5/3$ and 
$\nu=\pm 1/3$: while the former is an eigenstate of the SU(4)-symmetric Hamiltonian, the latter is not. It therefore requires an explicit spin-valley SU(4) symmetry breaking, 
\textit{e.g.} in the form of the above-mentioned Zeeman term $\Delta_Z$, to stabilize this state. However, a term as small as $\Delta_Z/(e^2/\epsilon l_B)\simeq 0.01$ is 
sufficient to stabilize the state (\cite{Papic2010}). Because $e^2/\epsilon l_B\simeq 109\sqrt{B[\text{T}]}$ K for graphene on h-BN and $\Delta_Z\simeq 1.2 B[\text{T}]$ K,
this condition is fulfilled for fields as small as $B\sim 1$ T, below the field values where the $1/3$-states have been observed experimentally (\cite{Dean2011,Polshyn2018}).

While numerical studies are less conclusive about their relevance at  $\nu=\pm 2/3$ and $\pm 4/3$ (\cite{Wu2015,Le2022}), Halperin wave functions may also be constructed for these filling factors, \textit{e.g.}
at $\nu=\pm 2/3$ with $m_1=1$, $m_2=m_3=m_4=3$, $n_{1\beta}=0$ and $n_{\alpha\beta}=3$ elsewise,
which amounts to having one spin-valley subbranch completely filled, while the electrons in the other three subbranches form a Laughlin state with an SU(3) symmmetry. As for the 
$\pm 1/3$-states, these states are not eigenstates of the SU(4)-symmetric Coulomb interaction and therefore require a (small) spin-valley degeneracy lifting. Similarly, the 
$\nu=\pm 4/3$ states may, in a first step, be approached in terms of a Lauglin state composed of electrons (or holes) in a single spin-valley subbranch, $m_1=3$, while the other intra-component exponents 
are $m_\alpha=1$, and there are no inter-component correlations, \textit{i.e.} $n_{\alpha\beta}=0$. From this perspective, the $\nu=\pm 4/3$ states are those that are closest 
to the one-component Laughlin states.

However, it is noteworthy to mention that there are other candidate states for $\nu=\pm 4/3$, and it is unlikely that the ground state be described 
in terms of these Halperin wave functions. Indeed, exact-diagonalization studies at $\nu=2/3-2=-4/3$ in graphene have shown that the state is a singlet in one of the channels 
(spin or valley) and fully polarized in the other one (\cite{AC2006,TJ2007,GR2007}), and the spectra show magnon-type modes. Notice that, due to the large number of components, 
one may construct other Halperin wave functions for this filling factor, as for example a state with $m_\alpha=3$ and $n_{\alpha\beta}=1$, or else one with 
$m_\alpha=3$, $n_{\alpha\beta}=n_{13}=n_{24}=3$ for the correlations inside each valley and $n_{12}=n_{34}=n_{14}=n_{23}=0$, \textit{i.e.} no correlations for electrons in different
valleys. While the latter trial state supports the numerically found magnon mode and has the correct ground-state degeneracy, there are no such modes in the former state in which the number
of electrons is fixed in all components. Finally, one needs to emphasize that the different states are in competition with one another and that explicit subordinate spin-valley 
symmetry-breaking terms become relevant for the choice between these states, as it has just recently been shown in exact-diagonalization studies on the $1/3$-family in graphene 
(\cite{Le2022}), and in some cases trial states are required to be constructed beyond the Halperin or multi-component CF states (\cite{Wu2015}).

\subsubsection{The particularity of the CF series $p/(2sp+1)$ in graphene}

The discussion of the $1/3$-FQHE states in graphene has shown to what extent the spin-valley degrees of freedom are relevant in understanding the effect. While one might 
think, in view of these results, that generally the system chooses a maximal spin-valley polarization compatible with the total filling factor, numerical studies of the CF series
at $\nu=p/(2sp+1)-2$ have shown that the situation is more involved (\cite{TJ2007}). From an electronic point of view, one may always find a CF wave function with complete 
spin-valley polarization for these filling factors, in which the CFs occupy $p$ CF LLs (\cite{Jain1989}), as mentioned in the introduction. This, however, does not seem to be the 
scenario chosen by Nature. Instead, at $p=2$ ($\nu=2/5 -2$), the electrons prefer to occupy another spin-valley branch instead of filling a second CF LL. This 
situation may either be described within an SU(4) spin-valley generalization of Jain's CF wave functions (\cite{TJ2007}) or in terms of the generalized Halperin wave function 
(\ref{eq:halperin}). It needs to be emphasized that both descriptions are equivalent for $p=1,...,4$.
Consider the situation of a small Zeeman effect that favours a spin over a valley polarization. In this case, the state at $\nu=2/5-2$ may be described by 
the wave function (\ref{eq:halperin}) with $m_\alpha=3$, $n_{\alpha\beta}=n_{13}=n_{24}=3$ for all inter-component correlations inside each valley and 
$n_{\alpha\beta}=n_{12}=n_{34}=n_{14}=n_{23}=2$. This state is a valley singlet, but the spin can be oriented freely for electrons inside each valley, \textit{i.e.} the 
state has a residual SU(2)$\times$SU(2) symmetry, and there are two spin-wave modes, one in each of the valleys. 

A particularly interesting situation arises at $\nu=4/9-2$, where all four components are filled such that the state is an SU(4) singlet with no magnon modes. 
This situation can be described in terms of the generalized Halperin wave function (\ref{eq:halperin}) with $m_\alpha=3$ and $n_{\alpha\beta}=2$ for all inter-component correlations,
and numerical calculations show that this unpolarized state is indeed favoured energetically over a completely polarized CF state (\cite{TJ2007}).

It is important to emphasize that such a completely unpolarized state or states with only partial polarization occur in this filling factor range of $-2<\nu<-1$ (and by particle-hole
symmetry in $1<\nu < 2$). This is not in line with common expectations from the often used single-particle picture, where one is tempted to successively populate the different spin-valley 
branches separated in energy by Zeeman-type terms. As mentioned above, these terms are much smaller than the leading Coulomb interaction scale, and it is the latter that needs to be considered
first. Also the picture of an exchange-driven maximal spin-valley polarization needs to be taken with care. While this picture is applicable at integer fillings such as 
$\nu=0$ and $\pm 1$ (and integer fillings different form the graphene IQHE series $\nu=\pm 2(2n+1)$), where the exchange energy generates the QHFM states, 
the numerical studies of the CF series $p/(2sp+1)$ in graphene indicate that
the special correlation energy built in the FQHE wave functions generally outcasts the exchange energy and thus gives rise to an unexpected non-monotonic behaviour of the 
spin-valley polarization if only the SU(4)-symmetric Coulomb repulsion is taken into account. However, as it has already been mentioned in the 
discussion of the $1/3$-family, the precise nature of the CF states realized in a typical experimental situation may sensitively depend on the explicit symmmetry-breaking terms. 
The energies of the different -- polarized or unpolarized -- trial states may be extremely close so that the latter subordinate terms are decisive in the choice of a particular
type of states even if they are not the relevant scale in the formation of these states (\cite{Abanin2013,Le2022}).

\subsubsection{Even-denominator states}

A particular FQHE state is the even-denominator state observed at $\nu=5/2$ and $7/2$ in 2D quantum-Hall systems in GaAs heterostructures (\cite{expMR}). 
From an experimental point of view, it seems settled now that they are spin-polarized (\cite{Piot}), as expected for the so-called Pfaffian state proposed by Moore and Read
(\cite{MR}). These states, however, observed in the $n=1$ LL require a particular form of the effective interaction potential that is in line with that (\ref{eq:effint}) for 
$n=1$ in the usual 2D electron system.\footnote{From a technical point of view, this can be shown within an expansion of the effective interaction potential in Haldane's 
pseudopotentials (\cite{HaldanePP}).} 
Right from the beginning of graphene research, it was realized that the particular form of the effective interaction potential (\ref{eq:effint}), if the appropriate form 
factors are taken into account, does not allow for the stabilization of the Moore-Read Pfaffian state (\cite{Goerbig2006}). However, it is possible to construct a multi-component
Halperin wave function also for this filling factor, \textit{e.g.} with $m_\alpha=3$ and again $n_{\alpha\beta}=n_{13}=n_{24}=3$ for all intercomponent exponents inside 
each valley and $n_{\alpha\beta}=n_{12}=n_{34}=n_{14}=n_{23}=1$. Alternatively, if a valley-degeneracy lifting outcasts the Zeeman effect, one may exchange the role of the 
spin and valley indices. Recent capacitance measurements have shown that even-denominator states at $\nu=\pm 1/2,\pm1/4$ can indeed be stabilized in the vicinity of a 
transition between two different spin-valley orders (\cite{Zibrov2018}), which gives credit to the multi-component interpretation of the effect (\cite{Roy2018})

\subsection{Electron crystals}\label{ssec:cryst}

While the FQHE states are certainly the most interesting correlated phases, due to their exotic quasiparticle excitations, they are, along with the QHFM phases, not the only 
correlated electronic phases that one encounters in partially filled LLs. Indeed, right after the discovery of the IQHE in the 1980ies, some researchers proposed a Wigner 
crystal to be the ground state in partially filled LLs, in order to minimize the repulsive Coulomb potential. This classical state is indeed realized at low filling factors, in which 
case the magnetic length becomes much smaller than the average electronic spacing so that quantum-mechanical effects due to the electronic wave-function overlap can be neglected. 
From an experimental point of view, it is nevertheless difficult to identify an electronic solid in conventional transport measurements: the electronic crystals are collectively 
pinned by the underlying impurities and therefore yield an insulating response in the same manner as an individual electronic localization does.

\begin{figure}[ht]
	\centering
		\includegraphics[width=0.68\textwidth]{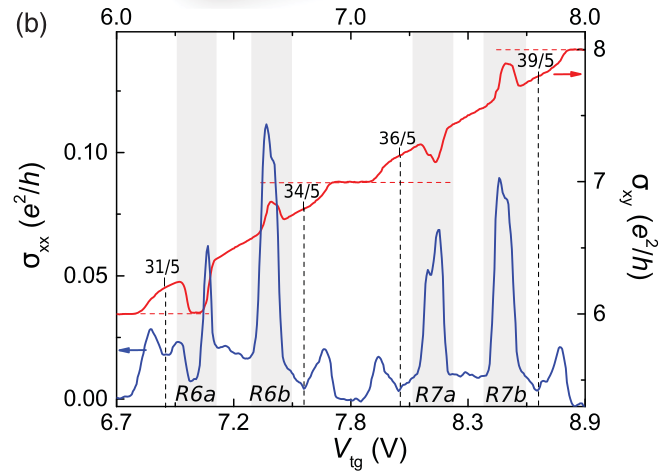}
	  \caption{Reentrant IQHE in graphene in the $n=2$ LL. The non-monotonic behaviour of the Hall conductance (red curve) as a function of the gate-voltage-induced electronic 
	  density is clearly visible at the four positions 
	  R6a, R6b, R7a, and R7b, where the conductances tend to their neighbouring integer values at $j=6$, 7 and 8 [adapted from (\cite{chen2019})]. }\label{fig06}
\end{figure}

From a theoretical point of view, electonic crystals become most interesting in higher LLs, where one does not only encounter the Wigner crystal but also more exotic bubble 
crystals, in which a lattice site is occupied by more than one electron, or stripe crystals. The phases have been theoretically predicted to arise in higher LLs in conventional 
GaAs heterostructures (\cite{FKS,moessner}), where \textit{e.g.} the stripe crystals yield a highly aniotropic electronic transport in the LLs $n=2$ (\cite{lilly99,du99}). 
One of the most salient features of electronic crystals in higher LLs is the so-called reentrant IQHE, which was discovered in GaAs heterostructures (\cite{cooper99,eisenstein02}).
The effect consists of a non-monotonic behaviour of the Hall resistance, in which a plateau corresponding to a FQHE is flanked by two plateaus at an integer value of $j$. 
The corresponding effect in graphene, which has been observed in 2019 (\cite{chen2019}), is shown in Fig. \ref{fig06}, where \textit{e.g.} the $31/5$ state (at $1/5$ filling 
of the $n=2$ LL) is flanked by two plateaus at $j=6$. While the plateau on the left-hand side can be understood in terms of single-particle localization of the few electrons 
occupying the lowest spin-valley branch of the $n=2$ LL, this is unlikely the case on the right-hand side (around $V_{tg}\simeq 7$ V). Theoretical calculations (\cite{knoester,chen2019})
indicate that the FQHE $1/5$ Laughlin state competes with a Wigner crystal that is extremely close in energy at this filling factor. Precisely at $\nu=31/5$, the FQHE state is lower 
in energy than the Wigner crystal, but once the filling factor is swept away from this precise value, the excited (gapped) quasiparticles or quasiholes raise the energy of the 
liquid states rapidly above that of the Wigner crystal, which becomes thus the ground state on both sides of the FQHE state and that gives rise to the insulating (IQHE) response there.

\section{Conclusions and Perspectives}


In conclusion, the observation of the FQHE and related correlated electronic phases in graphene corroborates the universality of the phenomenon: it is due to repulsive electronic
interactions between electrons inside a single partially filled LL, in which the electrons' kinetic energy is frozen. In addition to this flat-band scenario, which one may 
encounter also in other situations in condensed-matter systems, the chirality of the quantum states, inherited from the classical cyclotron motion with a specific sense imposed by the orientation of
the magnetic field, turns out to play an essential role in the formation of these exotic correlated states. The particularity of graphene is twofold: first, the states inherit 
the fourfold spin-valley degeneracy from the low-energy bands of graphene, and the associated SU(4) symmetry is to great extent respected by the leading Coulomb repulsion; second, the
specific spinorial form of the wave functions of graphene electrons yields slightly different wave-function overlaps as compared to the more conventional (GaAs) semiconductor 
heterostructures. Similarly, bilayer graphene is yet another version of a multi-component quantum-Hall system; in addition to the fourfold spin-valley degeneracy, the LLs $n=0$ and 
$n=1$ are situated at zero energy, and one is therefore confronted with an eightfold degeneracy here (\cite{Novoselov2006}). However, since the wave functions are different in the 
two LLs $n=0$ and $n=1$, the Coulomb interaction is not symmetric in this index so that there is no SU(8) symmetry as one might have expected at the beginning. Furthermore, the 
energy difference between the two layers and thus indirectly a \textit{valley Zeeman effect} can be tuned by an electric field applied perpendicular to the 2D system such that 
there is an interesting controllable tunability in this system, as it has been proposed theoretically (\cite{apalkov2010}) and measured experimentally (\cite{Maher2014}). In this framework it is also interesting to notice that the situation between monolayer and bilayer graphene can to some extent be interpolated in layered systems where two graphene monolayers are separated by an insulating h-BN film of variable thickness. While for large thickness ($d\gg l_B$) the monolayers are essentially decoupled, and one would expect simply to measure twice a monolayer quantum Hall effect, the interlayer coupling becomes stronger in the opposite limit ($d\ll l_B$) of thin h-BN layers. In the latter limit, novel interlayer-correlated FQHE states have indeed been observed experimentally (\cite{Liu2019}).

We have mainly explored, in this short review, the consequences of the SU(4) symmetry, which does not only impact in an unexpected manner the FQHE states in the form of 
highly non-monotonic spin-valley polarizations when increasing the filling factor in the zero-energy $n=0$ LL, but also the QHFM states that are formed at integer fillings 
beyond the particular graphene-IQHE sequence $\nu=\pm 2(2n+1)$. The QHFM phases, which are also relevant for many FQHE states that come along with certain forms of 
spin-valley polarization, host not only novel magnon types that are absent in conventional quantum-Hall systems with only the spin as an internal degree of freedom, but also a
rich zoo of exotic skrymions. While the theoretical classification of the QHFM states, as well as the spin-valley magnons and the skyrmions, is now rather complete, as a 
function of the subleading SU(4)-symmetry-breaking terms, the experimental investigation of these states has just started. As for the skyrmions, a promising first step has been 
made by scanning-tunneling-spectroscopic measurements that are capable of monitoring the particular sublattice occupation for valley-skyrmions with a spatially varying 
valley polarization (\cite{yazdani2022,Coissard2021}). In the case of magnons, a very interesting way of probing these collective excitations is that via non-local transport 
measurements with a high voltage bias between the source and the drain. While this non-local transport is expected to happen only at the edges (at low bias), it has been shown 
that at higher bias, above the Zeeman energy scale, magnons can be excited at the edges that propagate through the sample bulk and relax at another local gate tuned at 
$\nu=\pm 2$ (\cite{Stepanov2018,Wei2018,Assouline2021,Pierce2021}). This opens the perspective for a resistively-detected means of probing magnons in quantum-Hall systems,
namely in graphene. 

It is also interesting to mention that in line with the discovery of graphene as a particular 2D electronic system, other 2D crystals with unexpected eletronic porperties have 
been isolated. Most importantly, one should mention the family of 2D transition-metal dichalcogenides (\textit{e.g.} MoS$_2$, MoSe$_2$, WS$_2$ or WSe$_2$). Similarly to graphene,
the low-energy electronic properties of these materials may be described in terms of Dirac fermions, albeit \textit{massive} Dirac fermions now. Moreover, these materials show a 
significant spin-orbit interaction, most prominently in the valence band so that the spin-valley components are frozen there. Namely, the $n=0$ LL happens to occur in a single spin-valley 
component over a large magnetic-field and density range. In contrast to graphene, this allows for the study of the FQHE and other correlated phases in the ideal single-component 
limit for which the original theoretical trial wave functions (\cite{Laughlin1983,Jain1989,MR,GWW}) have been constructed. Recent compressibility measurements on high-quality WSe$_2$ 
samples have identified a large number of FQHE (\cite{Shi2020})
and are a promising path to more detailed studies of the FQHE in these materials, corroborating both the universality of the 
phenomenon in 2D electronic systems in a strong magnetic field and the specificities of the different systems in terms of multi- \textit{vs.} one-component manifestations of the 
effect.

\bibliographystyle{cas-model2-names}

\bibliography{refs}

\end{document}